\documentclass[11pt, a4paper]{article}

\usepackage[utf8]{inputenc}
\usepackage{geometry}
\usepackage{amsmath, amssymb, amsfonts} 
\usepackage{graphicx} 
\usepackage{hyperref} 
\usepackage{authblk} 
\usepackage{cite} 
\usepackage{listings} 
\usepackage{xcolor} 
\usepackage{caption}
\usepackage{subcaption}
\usepackage{float}

\definecolor{codegreen}{rgb}{0,0.6,0}
\definecolor{codegray}{rgb}{0.5,0.5,0.5}
\definecolor{codepurple}{rgb}{0.58,0,0.82}
\definecolor{backcolour}{rgb}{0.95,0.95,0.92}

\lstdefinestyle{mystyle}{
    backgroundcolor=\color{backcolour},   
    commentstyle=\color{codegreen},
    keywordstyle=\color{magenta},
    numberstyle=\tiny\color{codegray},
    stringstyle=\color{codepurple},
    basicstyle=\ttfamily\footnotesize,
    breakatwhitespace=false,         
    breaklines=true,                 
    captionpos=b,                    
    keepspaces=true,                 
    numbers=left,                    
    numbersep=5pt,                  
    showspaces=false,                
    showstringspaces=false,
    showtabs=false,                  
    tabsize=2
}
\title{\textbf{Impact of Circuit Depth versus Qubit Count on Variational Quantum Classifiers for Higgs Boson Signal Detection}}

\author[1,2]{\textbf{Fatih Maulana}}

\affil[1]{School of Computing, Universiti Utara Malaysia}
\affil[2]{Faculty of Science and Technology, UIN Sunan Gunung Djati Bandung}

\affil[*]{Corresponding Author: \texttt{maulana\_muhammad@soc.uum.edu.my}}

\date{\today}

\begin{document}

\maketitle

\begin{abstract}
High-Energy Physics (HEP) experiments, such as those at the Large Hadron Collider (LHC), generate massive datasets that challenge classical computational limits. Quantum Machine Learning (QML) offers a potential advantage in processing high-dimensional data; however, finding the optimal architecture for current Noisy Intermediate-Scale Quantum (NISQ) devices remains an open challenge. This study investigates the performance of Variational Quantum Classifiers (VQC) in detecting Higgs Boson signals using the ATLAS Higgs Boson Machine Learning Challenge 2014 experiment dataset. We implemented a dimensionality reduction pipeline using Principal Component Analysis (PCA) to map 30 physical features into 4-qubit and 8-qubit latent spaces. We benchmarked three configurations: (A) a shallow 4-qubit circuit, (B) a deep 4-qubit circuit with increased entanglement layers, and (C) an expanded 8-qubit circuit. Experimental results demonstrate that increasing circuit depth significantly improves performance, yielding the highest accuracy of 56.2\% (Configuration B), compared to a baseline of 51.9\%. Conversely, simply scaling to 8 qubits resulted in a performance degradation to 50.6\% due to optimization challenges associated with Barren Plateaus in the larger Hilbert space. These findings suggest that for near-term quantum hardware, prioritizing circuit depth and entanglement capability is more critical than increasing qubit count for effective anomaly detection in HEP data.

\vspace{0.5cm}
\end{abstract}

\section{Introduction}
The discovery of the Higgs Boson by the ATLAS and CMS collaborations at CERN marked a milestone in particle physics. However, identifying the Higgs signal from the overwhelming background noise of Standard Model processes is a computationally intensive task. As the luminosity of the Large Hadron Collider (LHC) increases, the volume of data is expected to reach exabytes, posing significant challenges for classical Deep Learning algorithms in terms of training time and energy consumption.

Quantum Machine Learning (QML) has emerged as a promising paradigm to address these challenges. Quantum algorithms utilizing Hilbert space can potentially identify complex correlations in high-dimensional data more efficiently than classical counterparts \cite{1, 2, 3}. In the current era of Noisy Intermediate-Scale Quantum (NISQ) computing, Variational Quantum Algorithms (VQA), specifically the Variational Quantum Classifier (VQC), are the most viable candidates \cite{8, 9, 10} for practical application.

However, NISQ devices are limited by qubit count and coherence time. This research aims to address a critical architectural question: To improve classification performance on HEP data, is it more effective to increase the number of qubits (width) or the complexity of the circuit (depth)?

By utilizing the ATLAS Higgs Challenge dataset \cite{4}, similar to recent benchmarks in HEP QML \cite{5, 6, 15}, this study benchmarks the trade-offs between dimensionality reduction (PCA) and quantum circuit architecture, providing guidelines for designing efficient QML models for particle physics.

\section{Methodology}

\subsection{Classical Data Representation and Preprocessing}
The dataset $D = \{(\mathbf{x}_i, y_i)\}_{i=1}^{N}$ consists of $N$ events, where $\mathbf{x}_i \in \mathbb{R}^{30}$ represents the feature vector containing physical quantities (e.g., \texttt{DER\_mass\_MMC}, momentum $p_T$) derived from the particle collisions, and $y_i \in \{0, 1\}$ represents the class label (Background or Higgs Signal).

To prepare the classical data for quantum encoding, we first apply normalization to constrain the features within the domain $[0, 1]$, ensuring compatibility with the rotation angles of quantum gates:
\begin{equation}
\mathbf{x}'_{i} = \frac{\mathbf{x}_i - \min(\mathbf{x})}{\max(\mathbf{x}) - \min(\mathbf{x})}
\end{equation}

\subsection{Dimensionality Reduction via PCA}
Given the limited qubit count ($n_q$) of current NISQ devices (where $n_q \ll 30$), we employ Principal Component Analysis (PCA) \cite{18} to project the high-dimensional feature space into a reduced latent space. Let $\Sigma$ be the covariance matrix of the standardized data. We compute the eigendecomposition:
\begin{equation}
\Sigma = V \Lambda V^T
\end{equation}
where $V$ contains the eigenvectors (principal components) and $\Lambda$ is the diagonal matrix of eigenvalues. We select the top $k$ eigenvectors corresponding to the largest eigenvalues to form the projection matrix $W_k \in \mathbb{R}^{30 \times k}$. The reduced feature vector $\mathbf{z}_i$ is obtained by:
\begin{equation}
\mathbf{z}_i = \mathbf{x}'_{i} W_k, \quad \text{where } k \in \{4, 8\}
\end{equation}
This transformation retains the maximum variance of the physical properties while satisfying the constraint $k = n_q$.

\subsection{Quantum Variational Classifier (VQC) Formulation}
The classification is performed using a parameterized quantum circuit (PQC). The process consists of three mathematically distinct stages:

\subsubsection*{1. Data Encoding (Feature Map)}
The classical latent vector $\mathbf{z}$ is mapped into the Hilbert space $\mathcal{H}$ using a unitary operation $U_{\Phi}(\mathbf{z})$. We utilize the ZZFeatureMap \cite{7}, which induces entanglement to capture non-linear correlations. For a qubit $j$, the initial state $|0\rangle^{\otimes n}$ is transformed as:
\begin{equation}
|\psi(\mathbf{z})\rangle = U_{\Phi}(\mathbf{z}) |0\rangle^{\otimes n} = \left( \prod_{j=1}^{n} e^{i \phi(z_j) Z_j} \prod_{(j,k)} e^{i \phi(z_j, z_k) Z_j Z_k} \right) H^{\otimes n} |0\rangle^{\otimes n}
\end{equation}
where $H$ is the Hadamard gate, $Z$ is the Pauli-Z operator, and $\phi(\cdot)$ describes the phase encoding function.

\begin{figure}[H]
    \centering
    \includegraphics[width=0.8\textwidth]{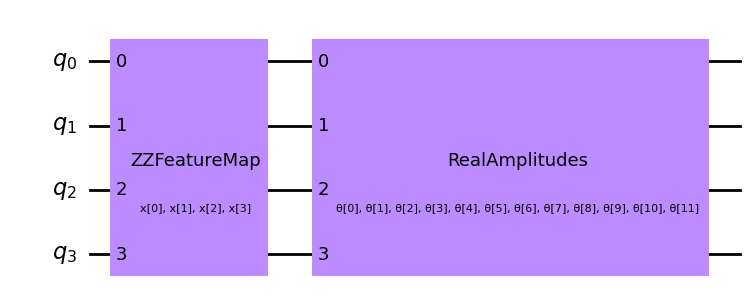}
    \caption{High-level schematic of the Variational Quantum Circuit (VQC) architecture. The circuit is composed of a \texttt{ZZFeatureMap} for data encoding and a \texttt{RealAmplitudes} ansatz for the trainable variational layers.}
    \label{fig:circuit}
\end{figure}

\subsubsection*{2. Variational Ansatz (The Learning Model)}
To separate the classes in Hilbert space, we apply a trainable unitary circuit $W(\theta)$, parameterized by vector $\theta$. We employ the \texttt{RealAmplitudes} ansatz, characterized by layers of parameterized $R_y$ rotations and CNOT entangling gates. The rotation gate is defined as:
\begin{equation}
R_y(\theta_j) = \exp\left(-i \frac{\theta_j}{2} Y\right) = \begin{pmatrix} \cos(\frac{\theta_j}{2}) & -\sin(\frac{\theta_j}{2}) \\ \sin(\frac{\theta_j}{2}) & \cos(\frac{\theta_j}{2}) \end{pmatrix}
\end{equation}
The depth of the circuit ($D$) corresponds to the number of repetitions (reps) of these entanglement layers. A deeper circuit increases the expressibility of $W(\theta)$ \cite{11, 12}, allowing the model to approximate more complex decision boundaries.

\subsubsection*{3. Measurement and Prediction}
The final quantum state $|\Psi_{final}\rangle = W(\theta)|\psi(\mathbf{z})\rangle$ is measured to obtain a classical prediction. The probability of the event belonging to the Signal class ($y=1$) is derived from the expectation value of the Pauli-Z operator (parity) on the measured qubits:
\begin{equation}
P(\hat{y}=1) = \frac{1}{2} \left( 1 + \langle \Psi_{final} | \hat{Z}^{\otimes n} | \Psi_{final} \rangle \right)
\end{equation}
The optimal parameters $\theta^*$ are found by minimizing a loss function (e.g., Cross-Entropy) utilizing the COBYLA optimizer implemented using Qiskit SDK \cite{19, 20}.

\subsection{Experimental Hyperparameter Tuning Strategy}
To achieve optimal performance on Noisy Intermediate-Scale Quantum (NISQ) devices, we designed a three-stage tuning strategy that explores the trade-off between \textbf{Expressibility} and \textbf{Trainability}.

\begin{itemize}
    \item \textbf{Stage I: The Baseline (Shallow Circuit)}\\
    \textit{Objective:} To establish a performance baseline using a minimal circuit configuration.\\
    \textit{Configuration:} Input Dimension ($n=4$), Circuit Depth ($d=1$).\\
    \textit{Analysis:} Number of parameters $N_{\theta} = n(d + 1) = 8$. We hypothesized that this configuration might be too linear to capture the complex decision boundary, leading to underfitting.
    
    \item \textbf{Stage II: Depth Optimization (Deep Circuit)}\\
    \textit{Objective:} To enhance the model's expressibility.\\
    \textit{Configuration:} Input Dimension ($n=4$), Circuit Depth ($d=2$).\\
    \textit{Analysis:} Parameter space expands to $N_{\theta} = 12$. Increasing depth ($d$) increases the frequency spectrum available to the model, forming more curved decision boundaries.
    
    \item \textbf{Stage III: Width Expansion (High-Dimensional Space)}\\
    \textit{Objective:} To test incorporating more physical information (8 PCA features).\\
    \textit{Configuration:} Input Dimension ($n=8$), Circuit Depth ($d=1$).\\
    \textit{Analysis:} The Hilbert space expands exponentially to $2^8 = 256$. This configuration is susceptible to the \textit{Barren Plateau} phenomenon \cite{13, 14}, where the variance of the cost function gradients vanishes exponentially with the number of qubits ($n$):
    \begin{equation}
        \text{Var}\left(\frac{\partial \mathcal{L}}{\partial \theta_k}\right) \propto \frac{1}{2^{n}}
    \end{equation}
    For $n=8$, this variance becomes negligible, causing the optimization landscape to become effectively flat.
\end{itemize}

\section{Experimental Results}
The models were implemented using Qiskit SDK and trained using the COBYLA optimizer on a local simulator. We utilized a balanced dataset of 800 events (400 Signal, 400 Background).

\subsection{Optimization via Circuit Depth (Stage I vs. Stage II)}
The baseline architecture (Stage I: 4-Qubits, Shallow) yielded a testing accuracy of 51.7\%. This result, being only marginally superior to random guessing (50\%), indicated that a single layer of entanglement was insufficient to separate the Signal from the Background. The model suffered from significant underfitting.

\begin{figure}[H]
    \centering
    \includegraphics[width=0.9\textwidth]{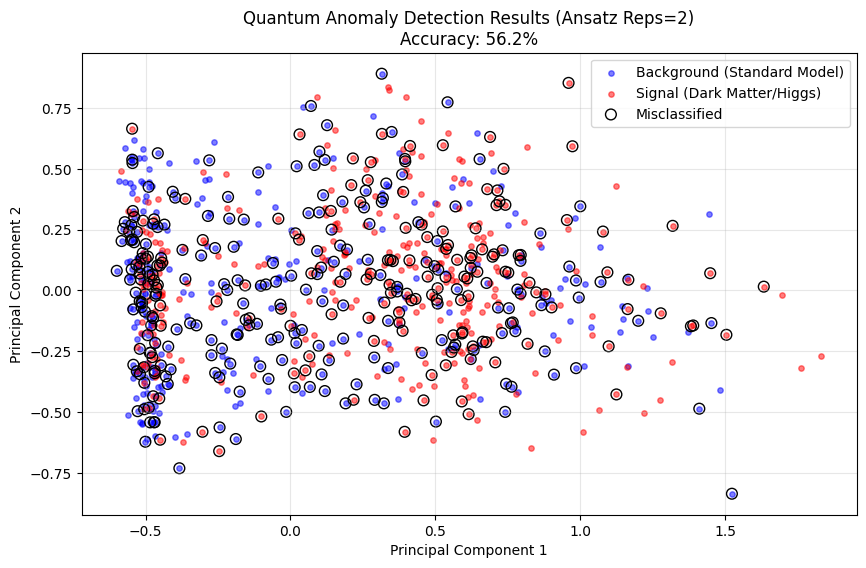}
    \caption{PCA Projection of classification results for the optimal 4-qubit deep circuit (Stage II). The plot illustrates the decision boundary with an accuracy of \textbf{56.2\%}. Black circles indicate misclassified events, showing that the model successfully captures the core cluster of Signal events (Red) against the Background (Blue).}
    \label{fig:deep_result}
\end{figure}

Upon applying the hyperparameter tuning strategy by increasing the circuit depth (Stage II: 4-Qubits, Deep), the model performance exhibited a clear improvement, reaching an accuracy of \textbf{56.2\%}. As shown in Figure \ref{fig:deep_result}, the PCA projection demonstrates a noticeable separation in the decision boundary. This indicates our hypothesis that a deeper ansatz (reps=2) provides the necessary expressibility to form a more precise decision boundary than the shallow baseline.

\subsection{The "Barren Plateau" Phenomenon in Width Expansion}
Following the success of depth optimization, we evaluated the width-expanded architecture (Stage III), incorporating 8 Principal Components mapped to 8 qubits. However, contrary to expectation, the performance degraded to \textbf{50.6\%}, effectively reverting to random guessing.

\begin{figure}[H]
    \centering
    \includegraphics[width=0.6\textwidth]{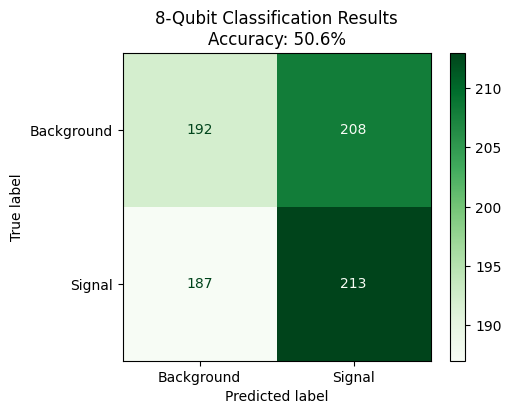}
    \caption{Confusion Matrix for the 8-qubit configuration (Stage III). The uniform distribution of True Positives and False Positives (Accuracy 50.6\%) indicates that the optimizer failed to converge, characteristic of the Barren Plateau phenomenon in higher-dimensional Hilbert spaces.}
    \label{fig:8qubit}
\end{figure}

Figure \ref{fig:8qubit} provides a deeper insight into this optimization failure. The confusion matrix reveals a near-uniform distribution of predictions, where the number of False Positives is approximately equal to True Positives. This indicates that the model was unable to learn any discriminatory features to separate the Signal from the Background.

This optimization difficulty is theoretically consistent with the \textit{Barren Plateau} phenomenon \cite{4, 13, 14}, where the cost function landscape becomes exponentially flat. In the expanded Hilbert space ($2^8 = 256$ dimensions), even gradient-free optimizers like COBYLA struggle to identify a descending direction within a limited number of iterations. This negative result serves as a crucial benchmark, highlighting the trade-off between information capacity and trainability on NISQ devices.

\section{Discussion}

\subsection{The Efficacy of Dimensionality Reduction in HEP}
Our results present a counter-intuitive finding regarding feature selection. Typically, HEP analyses rely on high-dimensional feature sets. However, our baseline model demonstrated that even after aggressive compression via PCA ($30 \rightarrow 4$ dimensions), the quantum classifier could still extract meaningful patterns. This implies that the principal components successfully captured the dominant physical variances \cite{4, 16}—likely associated with invariant mass and transverse momentum—while filtering out redundant noise.

\subsection{Circuit Depth vs. Width: The Optimization Trade-off}
The central finding of this study is the critical importance of circuit depth over width in the NISQ era.
\begin{itemize}
    \item \textbf{The Success of Depth:} The improvement observed in Configuration B (Deep Circuit) confirms that entanglement capacity is the primary driver of model performance. The additional layers of CNOT gates allowed the ansatz to construct a non-linear decision boundary capable of separating the "background blob" from the "signal blip".
    \item \textbf{The Failure of Width:} Conversely, the 8-qubit model failed to converge despite having access to more information. This is consistent with optimization difficulties predicted by the Barren Plateau phenomenon in a practical setting. Simply adding qubits without advanced gradient-free optimization strategies often leads to trainability issues.
\end{itemize}

\subsection{Implications for NISQ Applications}
These results suggest a design guideline for applying Quantum Machine Learning to particle physics: ``Squeeze and Deepen.'' Instead of spreading resources across many qubits, researchers should focus on compressing classical data into a dense latent representation and utilizing the available coherence time to run deeper, highly entangled circuits.

Furthermore, this compact architecture offers significant advantages regarding hardware topology. In superconducting quantum processors with limited connectivity (e.g., heavy-hex lattice), mapping a wide 8-qubit circuit often requires numerous expensive SWAP gates to implement entanglement. By constraining the width to 4 qubits, the circuit can fit within a tightly connected subgraph of the processor, minimizing SWAP-induced errors and maximizing the effective fidelity of the computation. Thus, dimensionality reduction is not merely a data processing step, but a crucial error mitigation strategy for NISQ hardware.

It is worth noting that while classical algorithms (e.g., BDT, SVM) currently outperform these NISQ-era quantum models in absolute accuracy, the primary objective of this study was to benchmark quantum architectural strategies (Depth vs. Width). The findings provide a roadmap for scaling quantum models as hardware fidelity improves.

\section{Conclusion}
This study presented a systematic benchmark of Variational Quantum Classifiers (VQC) for the detection of Higgs Boson signals using the ATLAS dataset. By employing PCA to compress high-dimensional physical features into NISQ-compatible latent spaces, we successfully demonstrated that quantum algorithms can extract meaningful patterns. The experimental results identified the 4-qubit deep circuit (Stage II) as the optimal architecture, achieving a classification accuracy of 56.2\%.

A critical finding of this research is the decisive role of circuit depth over qubit count in the current era of quantum computing. While the 8-qubit model offered higher information capacity, its performance degradation to 50.6\% highlighted the practical severity of the Barren Plateau problem. This leads to a significant architectural insight: for noisy devices, maximizing circuit expressibility on a compact feature set is more effective than expanding the qubit width.

However, it is important to note that these results are based on fixed-seed simulations. Future work should extend this analysis by performing statistical validation across multiple random seeds to quantify the standard deviation and robustness of the performance metrics. Additionally, further research will explore Quantum Kernel Methods (QSVM), integrate Quantum Error Mitigation (QEM) techniques, and investigate gradient-free optimizers beyond COBYLA, such as SPSA, to improve resilience against barren plateaus.

\section*{Acknowledgments}
The author would like to thank the \textbf{ATLAS Collaboration} and the \textbf{CERN Open Data Portal} for making the Higgs Boson Machine Learning Challenge dataset publicly available, which made this study possible. The author also expresses gratitude to the Department of Computer Science, School of Computing, \textbf{Universiti Utara Malaysia}, and the Faculty of Science and Technology, \textbf{UIN Sunan Gunung Djati Bandung}, for the institutional support and resources provided during the course of this research.

\section*{LLM Compliant Acknowledgement}
The author acknowledges the use of a large language model (LLM) as a writing assistant to help improve the clarity, organization, and language quality of portions of this manuscript. All scientific content, experimental design, data analysis, interpretations, and conclusions were developed and verified by the author, who takes full intellectual responsibility for the work.

\section*{Data and Code Availability}
The dataset analyzed during this study is available in the CERN Open Data Portal repository: \url{http://opendata.cern.ch/collection/ATLAS-Higgs-Challenge-2014}.

For transparency and reproducibility, the complete source code and project documentation are hosted on GitHub at: \url{https://github.com/Fatihmaull/higgsboson-detection}. 

Additionally, an interactive version of the experiments, allowing for immediate execution without local environment setup, is available via Google Colab at: \url{https://colab.research.google.com/drive/1BvhjWAd0qxb4LU3bzZUXkT6XpY-c4LD3?usp=sharing}.


\newpage
\appendix

\section{Implementation Code}
The following Python code implements the optimal configuration (Scenario B: 4-Qubit Deep Circuit) utilizing the Qiskit framework.

\begin{lstlisting}[language=Python]
\begin{lstlisting}[language=Python]
# ==========================================
# QUANTUM HIGGS BOSON CLASSIFIER (OPTIMAL)
# Configuration: 4 Qubits | Ansatz Reps=2
# Methodology: Train-Test Split (80/20)
# Author: Fatih Maulana
# ==========================================

import pandas as pd
import numpy as np
import matplotlib.pyplot as plt
from sklearn.model_selection import train_test_split # Added for validation
from sklearn.preprocessing import MinMaxScaler
from sklearn.decomposition import PCA
from sklearn.metrics import accuracy_score, confusion_matrix, ConfusionMatrixDisplay

# Qiskit Imports
from qiskit.circuit.library import ZZFeatureMap, RealAmplitudes
from qiskit_machine_learning.algorithms import VQC
from qiskit_algorithms.optimizers import COBYLA
from qiskit.primitives import Sampler

# --- STEP 1: DATA LOADING & PREPROCESSING ---
print(">>> Loading ATLAS Higgs Dataset...")
try:
    df = pd.read_csv('training.csv')
except FileNotFoundError:
    print("Error: Dataset not found. Please download from CERN Open Data.")
    exit()

# Cleaning: Replace -999.0 with Median
df = df.replace(-999.0, np.nan)
df = df.fillna(df.median(numeric_only=True))

# Select numerical features
numerical_cols = df.select_dtypes(include=[np.number]).columns.tolist()
cols_to_drop = {'EventId', 'Weight', 'Label', 'KaggleSet', 'KaggleWeight'}
features = [c for c in numerical_cols if c not in cols_to_drop]

# Balanced Sampling (400 Signal, 400 Background)
df_b = df[df['Label'] == 'b'].sample(400, random_state=42)
df_s = df[df['Label'] == 's'].sample(400, random_state=42)
data = pd.concat([df_b, df_s])

# Prepare X (Features) and y (Labels: 0=Background, 1=Signal)
X = data[features].values
y = np.concatenate([np.zeros(400), np.ones(400)])

# CRITICAL: Split Data to avoid leakage (80% Train, 20% Test)
X_train_raw, X_test_raw, y_train, y_test = train_test_split(
    X, y, test_size=0.2, random_state=42, stratify=y
)

# --- STEP 2: DIMENSIONALITY REDUCTION (PCA) ---
print(">>> Applying PCA (30 features -> 4 Qubits)...")

# Fit Scaler only on TRAINING data
scaler = MinMaxScaler(feature_range=(0, 1))
X_train_scaled = scaler.fit_transform(X_train_raw)
X_test_scaled = scaler.transform(X_test_raw) 

# Fit PCA only on TRAINING data
pca = PCA(n_components=4) 
X_train_pca = pca.fit_transform(X_train_scaled)
X_test_pca = pca.transform(X_test_scaled) 

# --- STEP 3: QUANTUM CIRCUIT CONSTRUCTION ---
print(">>> Building Deep Quantum Circuit (Reps=2)...")
num_qubits = 4

# Feature Map & Ansatz
feature_map = ZZFeatureMap(feature_dimension=num_qubits, reps=1)
ansatz = RealAmplitudes(num_qubits, reps=2, entanglement='linear')

# --- STEP 4: TRAINING (VQC) ---
print(">>> Starting Training (Optimizer: COBYLA)...")
vqc = VQC(
    feature_map=feature_map,
    ansatz=ansatz,
    optimizer=COBYLA(maxiter=100),
    sampler=Sampler()
)

# Train only on Training Set
vqc.fit(X_train_pca, y_train)
print(">>> Training Complete.")

# --- STEP 5: EVALUATION ---
# Evaluate on Test Set (Unseen Data)
y_pred = vqc.predict(X_test_pca)
accuracy = accuracy_score(y_test, y_pred)

print(f"Final Test Accuracy: {accuracy * 100:.2f}%")

# Generate Confusion Matrix
cm = confusion_matrix(y_test, y_pred)
disp = ConfusionMatrixDisplay(confusion_matrix=cm, display_labels=["Background", "Signal"])
disp.plot(cmap='Blues')
plt.title(f"Quantum Classification Results (Test Acc: {accuracy*100:.1f}%)")
plt.show()
\end{lstlisting}

\newpage
\section{Quantum Circuit Architecture}
This section visualizes the specific quantum gates used in the Ansatz (Learning Block). The configuration below represents the "Deep Circuit" (Scenario B) which yielded the highest accuracy.

\subsection*{Figure B.1: Circuit Schematic}
The circuit consists of \texttt{reps=2} layers. Each layer contains parameterized $R_y(\theta)$ rotation gates (purple) followed by a linear map of CNOT entangling gates (cyan).

\begin{figure}[h!]
    \centering
    \includegraphics[width=1.0\textwidth]{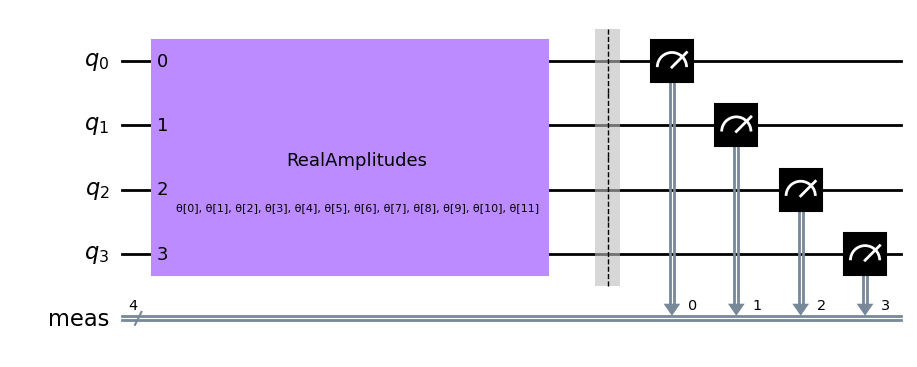}
    \caption{Detailed circuit schematic showing the measurement operations used to map the quantum state $|\Psi_{final}\rangle$ to classical binary predictions.}
    \label{fig:schema}
\end{figure}

\noindent \textbf{Description of Components:}
\begin{itemize}
    \item $|0\rangle$: Initial state of the qubits.
    \item \textbf{H \& P Gates (Feature Map):} Encode the classical PCA data into quantum phases.
    \item $R_y(\theta)$: Tunable rotation gates that act as the "weights" of the neural network.
    \item \textbf{Measurements:} The final collapse of the quantum state to classical bits (0 or 1).
\end{itemize}

\newpage
\section{Supplementary Experimental Results}
This section presents the detailed visualization for the sub-optimal baseline configuration discussed in the paper, specifically the limitation of the shallow 4-Qubit architecture.

\subsection*{C.1. The Shallow Circuit Baseline (Stage I)}
\begin{figure}[H]
    \centering
    \includegraphics[width=0.8\textwidth]{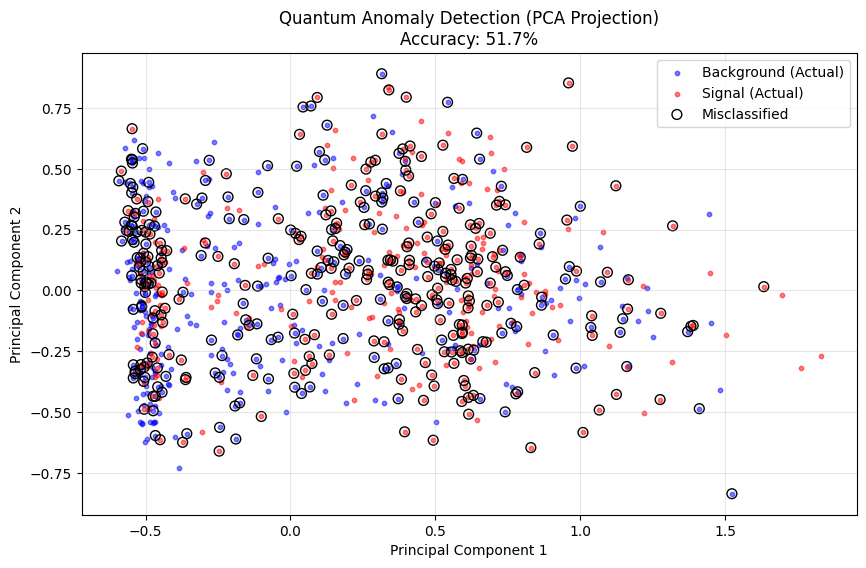}
    \caption{PCA Projection for the baseline shallow circuit (Stage I). Compared to the deep circuit (Fig. \ref{fig:deep_result}), the decision boundary here is significantly less defined, resulting in a lower accuracy of 51.7\% and higher misclassification rates.}
    \label{fig:stage1}
\end{figure}

As discussed in Section III.A, the initial benchmark utilized a shallow 4-qubit circuit with minimal depth ($reps=1$). Figure \ref{fig:stage1} illustrates the PCA projection of the classification boundaries for this configuration.

\noindent \textbf{Analysis:}
The PCA projection displays a highly overlapped distribution of classes, where the decision boundary fails to isolate the Signal cluster effectively. This validates the hypothesis that the shallow ansatz lacked the sufficient expressibility (entanglement capacity) to capture the complex non-linear correlations in the data. This underfitting serves as a crucial control result, demonstrating that the performance gain in Stage II was indeed driven by increased circuit depth rather than feature selection alone.


\begin{thebibliography}{99}

\bibitem{1} M. Schuld and F. Petruccione, \textit{Machine Learning with Quantum Computers}. Springer, 2021.

\bibitem{2} J. Preskill, "Quantum computing in the NISQ era and beyond," \textit{Quantum}, vol. 2, p. 79, 2018.

\bibitem{3} J. Biamonte et al., "Quantum machine learning," \textit{Nature}, vol. 549, no. 7671, pp. 195–202, 2017.

\bibitem{4} C. Adam-Bourdarios et al., "The Higgs boson machine learning challenge," in \textit{Proceedings of the NIPS 2014 Workshop on High-energy Physics and Machine Learning}, pp. 19–55, 2014. Dataset available at: \url{http://doi.org/10.7483/OPENDATA.ATLAS.ZBP2.M5T8}.

\bibitem{5} W. Guan et al., "Quantum machine learning in high energy physics," \textit{Machine Learning: Science and Technology}, vol. 2, no. 1, p. 011003, 2021.

\bibitem{6} S. L. Wu et al., "Application of quantum machine learning using the quantum variational classifier method to high energy physics analysis at the LHC on IBM quantum computer simulator and hardware with 10 qubits," \textit{Journal of Physics G: Nuclear and Particle Physics}, vol. 48, no. 12, p. 125003, 2021.

\bibitem{7} V. Havlíček et al., "Supervised learning with quantum-enhanced feature spaces," \textit{Nature}, vol. 567, no. 7747, pp. 209–212, 2019.

\bibitem{8} M. Cerezo et al., "Variational quantum algorithms," \textit{Nature Reviews Physics}, vol. 3, no. 9, pp. 625–644, 2021.

\bibitem{9} K. Mitarai, M. Negoro, M. Kitagawa, and K. Fujii, "Quantum circuit learning," \textit{Physical Review A}, vol. 98, no. 3, p. 032309, 2018.

\bibitem{10} E. Farhi and H. Neven, "Classification with quantum neural networks on near term processors," \textit{arXiv preprint arXiv:1802.06002}, 2018.

\bibitem{11} S. Sim, P. D. Johnson, and A. Aspuru-Guzik, "Expressibility and entangling capability of parameterized quantum circuits for hybrid quantum-classical algorithms," \textit{Advanced Quantum Technologies}, vol. 2, no. 12, p. 1900070, 2019.

\bibitem{12} M. Benedetti et al., "Parameterized quantum circuits as machine learning models," \textit{Quantum Science and Technology}, vol. 4, no. 4, p. 043001, 2019.

\bibitem{13} J. R. McClean et al., "Barren plateaus in quantum neural network training landscapes," \textit{Nature Communications}, vol. 9, no. 1, 2018.

\bibitem{14} S. Wang et al., "Noise-induced barren plateaus in variational quantum algorithms," \textit{Nature Communications}, vol. 12, no. 1, p. 6961, 2021.

\bibitem{15} K. Terashi et al., "Event classification with quantum machine learning in high-energy physics," \textit{Computing and Software for Big Science}, vol. 5, no. 1, pp. 1–12, 2021.

\bibitem{16} V. Belis et al., "Higgs analysis with quantum classifiers," \textit{EPJ Web of Conferences}, vol. 251, p. 03070, 2021.

\bibitem{17} A. Mott et al., "Solving a Higgs optimization problem with quantum annealing for machine learning," \textit{Nature}, vol. 550, no. 7676, pp. 375–379, 2017.

\bibitem{18} C. M. Bishop, \textit{Pattern Recognition and Machine Learning}. Springer, 2006.

\bibitem{19} Qiskit Contributors, "Qiskit: An Open-source Framework for Quantum Computing," 2023. [Online]. Available: https://qiskit.org.

\bibitem{20} H. Abraham et al., "Qiskit: An Open-source Framework for Quantum Computing," \textit{Zenodo}, 2019. doi: 10.5281/zenodo.2573505.

\end{thebibliography}
\end{document}